\documentclass[aps,prb,manuscript]{revtex4}
\usepackage{graphicx}
\usepackage{epsfig}
\usepackage{color}
\usepackage{bm}

\begin{document}
\draft
\title {Diverse quantization phenomena in AA bilayer silicene}

\author{Po-Hsin Shih$^{1}$, Thi-Nga Do$^{1}$, Godfrey Gumbs$^{2,3}$, Danhong Huang$^{4}$, Hai Duong Pham$^{1}$, and Ming-Fa Lin$^{1}$}
\affiliation{$^{1}$Department of Physics, National Cheng Kung University, Tainan, Taiwan 701\\
$^{2}$Department of Physics and Astronomy, Hunter College of the City University of New York,
 695 Park Avenue, New York, New York 10065, USA\\
$^{3}$Donostia International Physics Center (DIPC), P de Manuel Lardizabal,
 4, 20018 San Sebastian, Basque Country, Spain\\
$^{4}$US Air Force Research Laboratory, Space Vehicles Directorate, Kirtland Air Force Base,
New Mexico 87117, USA}

\date{\today}

\begin{abstract}
The rich magneto-electronic properties of AA-bottom-top (bt) bilayer silicene are investigated using a generalized tight-binding model. The electronic structure exhibits two pairs of oscillatory energy bands in which the lowest conduction and highest valence states of the low-lying pair are away from the K point. The quantized Landau levels (LLs) are classified into various separated groups by the localization behaviors of spatial distributions. The LLs in the vicinity of Fermi energy do not present simple wave function modes which are quite different from other two-dimensional systems. The geometry symmetry, intralayer and interlayer atomic interactions, and effect of a perpendicular magnetic field are responsible for the peculiar LL energy spectra in AA-bt bilayer silicene. This work provides a better understanding of the diverse magnetic quantization phenomena in 2D condensed-matter materials.

\end{abstract}
\pacs{PACS:}
\maketitle

\section{Introduction}

Silicene, an isostructure to graphene, is purely made of silicon atoms through both the ${sp^2}$ and ${sp^3}$ bondings.
Up to this point of time, silicene systems have been successfully synthesized by the epitaxial growth on the different substrate surfaces.
Monolayer silicene with different sizes of the unit cells was produced on various substrates, such as Ag(111) (${4\times\,4}$) \cite{1, 2}, Ir(111) (${\sqrt 3\times\,\sqrt 3}$) \cite{3} and ZrB$_2$(0001) (${2\times\,2}$) \cite{4}.
Such a buckled single-layer honeycomb lattice was clearly identified by the high-resolution measurements of STM and low-energy electron diffraction. It should be noticed that the examined system might present an enlarged unit cell because of the significant effect of substrates, e.g., the non-negligible orbital hybridization between silicon atoms and substrate ones.
The energy band shows a Dirac-cone structure with a graphene-like Fermi velocity but in the presence of a small spin-orbital coupling of $\sim\,,2$ meV, as directly confirmed by the ARPES experiments \cite{1, 2}.
Bilayer silicene is predicted to exhibit four distinct kinds of stacking configurations, namely AB-bt, AB-bb, AA-bt, and AA-bb \cite{5, 6, 7, 8, 9, 10, 11, 12}. Experimental growth has been done for the two AB stable bucklings \cite{5}, however, there is still lack of similar experimental study for AA-stacked bilayer silicene.\\

Theoretical studies have focused on the essential properties of monolayer and bilayer with or without adatom chemisorptions or guest atom substitutions based on various approaches, covering the first-principles calculations \cite{ 6, 7, 8, 9, 10, 11, 12}, the generalized tight-binding model \cite{13, 14}, and the effective-mass approximation \cite{15}. While the first method is suitable in studying the optimal geometries, the later two are powerful tools for the exploration of the magnetic quantizations. Nevertheless, the effective-mass model reveals certain limitations in dealing with complicated systems such as ABC trilayer graphene \cite{ABC}, AB-bt bilayer silicene \cite{ABbt} and others. 
A single-layer silicene is predicted to possesses a relatively narrow gap of ${\sim\,5}$ meV \cite{6}, as a result of the weak spin-orbital coupling.
The interplay between intrinsic interactions and electric or magnetic fields might cause the destruction of ${z=0}$-plane mirror symmetry (electric field) and the appearance of periodical Peierles phases (magnetic field), respectively. The former gives rise to the spin-dominated splitting energy band and the dramatic change of band gap \cite{15}; the latter yields highly degenerate Landau levels \cite{15}.
Recently, silicene with adatoms (X) systems have been investigated for the chemical modifications by the remarkable multi-orbital hybridizations in Si-Si, Si-X and X-X bonds by VASP calculations \cite{16, 17}. Such phenomena might induce the metallic or semiconducting behaviors in materials.
However for bilayer silicene, the phenomenological models might not be suitable for solving the magnetic-field-dominated fundamental properties due to the non-negligible bucklings, the largely enhanced spin-orbital interactions and the complex interlayer hopping integrals.
Specifically, the optimization of reliable tight-binding parameters which are necessary to reproduce the low-lying two pairs of valence and conduction bands are likely impossible because of the complex in the first-principle results for energy dispersions \cite{6, 7, 8, 9, 10, 11, 12}. Bilayer silicene with AA and AB stackings are predicted to present the non-monotonous energy dispersions and the irregular valleys at the non-high-symmetry points. Moreover, the free carrier densities are expected to be quite sensitive to the buckling and stacking configurations. \\

In this paper, we explore the diverse quantization phenomena in AA-bt silicene by employing a generalized tight-binding model. Calculations and analysis will focus on the band properties across the Fermi level, energy dispersions, critical points in energy-wave-vector space, distinct valleys, special structures of van Hove singularities in density of states, significant magnetic-field dependences, classification of valence and conduction LL groups and their main features. Especially, the valley-enriched magnetic quantization will be investigated by the detailed examinations on the spatial oscillation modes of magnetic wave functions. The interesting combined effects of distinct LL groups are clarified from different stable or metastable valleys. Furthermore, we also discuss the significant differences in the essential physical properties between AA-bt bilayer silicene and monolayer silicene and graphene in the electronic valley structure point of view. \\

\section{Theoretical Method}

We develop a generalized tight-binding model for AA-bt bilayer silicene in the presence of a uniform perpendicular magnetic field and explore the magnetoelectronic properties. The crystal structure of a AA-bt bilayer silicene with hopping interaction terms is clearly illustrated in Fig. 1(a). The two layers possess opposite buckling order, in which the [A$^1$, A$^2$] sublattices lie in the inner planes while the [B$^1$, B$^2$] sublattices are located at the outer planes. The lattice constant and bond length are $a=3.83$ $\AA$ and $b=2.21$ $\AA$, respectively. A primitive unit cell contain four silicon atoms. Accordingly, the critial Hamiltonian is built from the four tight-binding functions of Si-3p$_z$ orbitals. It can be written as

\begin{equation}
\nonumber
H=\sum_{i,l}U^{l}_{i}c^{\dagger l}_{i}c^{l}_{i}+\sum_{\langle i,j \rangle,l,l'}\gamma^{l l'}_{i j}c^{\dagger l}_{i}c^{l'}_{j}.
\end{equation}

Here, $c^{\dagger l}_{i}$ (${c^l_i}$) is the creation (annihilation) operator which could generate (destroy) an electronic state at the $i$-th site of the $l$-th layer, ${U^l_i(A^l, B^l)}$ is the buckled-sublattice-height-dependent Coulomb potential energy due to the applied gate voltage. $\gamma^{l l'}_{i j}$ represents the intra- and inter-layer atomic interactions, in which the former comes from the the nearest-neighbor interactions in the [${A^l, B^l}$] sublattices (${\gamma_0\,=1.004}$ eV) and the latter stand for interactions between sublattices from different layers (${\gamma_1\,=-2.110}$ eV, ${\gamma_2\,=-1.041}$ eV, and ${\gamma_3\,=0.035}$ eV), as shown in Fig. 1(a).
Interestingly, our optimization show that $\gamma_1$ is much higher than while $\gamma_2$ is comparable to $\gamma_0$. This result, in addition to the high symmetry of stacking configuration, might be responsible for the negligible spin-orbital couplings. Apparently, the above Hamiltonian, in the absence of spin-orbital couplings, is sufficient for the investigation of certain essential physical properties.  \\

\begin{figure}
\centering
{\includegraphics[width=0.8\linewidth]{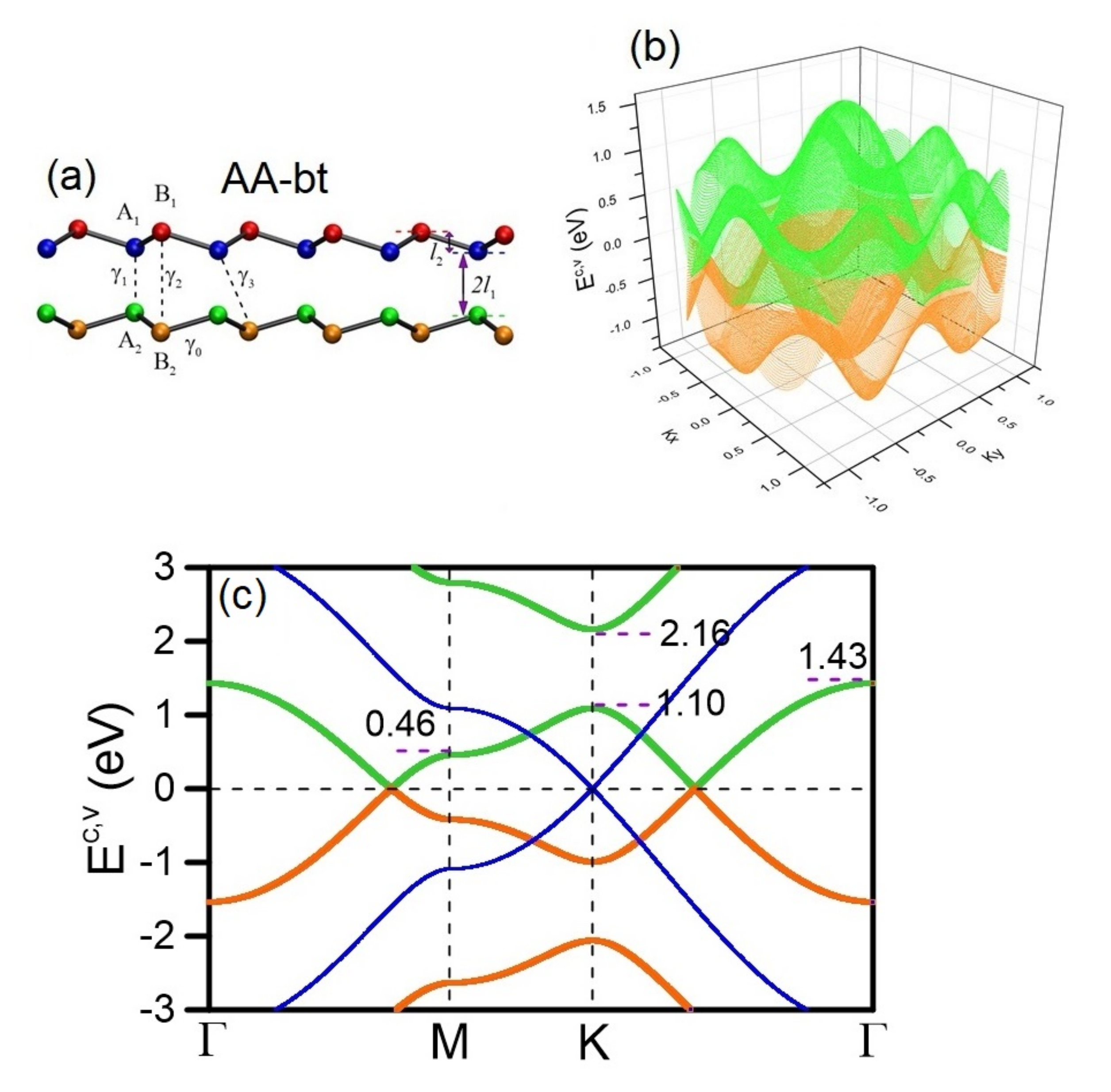}}
\caption{(Color online) The (a) side view of geometric structure including the significant hopping integrals for AA-bt bilayer silicene. The low-energy band structures are shown for (b) 3D and (c) 2D along the high symmetry points. Energy band of monolayer silicene is also plotted in (c) for comparison.}
\label{Figure 1}
\end{figure}

The application of a uniform perpendicular magnetic field evidently changes the main characteristic of lattice. The original unit cell is remarkably enlarged to be come a long rectangular due to the field-induced extra Peierls phases \cite{ABbt}. The extended unit cell includes 16$\phi_0 /\phi$ Si atoms, in which $\phi_0 =hc/e$ is the magnetic flux
quantum and $\phi = B_z \sqrt{3}a^2/2$ is the magnetic flux. As a result, the magnetic Hamiltonian is a huge Hermitian matrix, e.g., the size is a ${\sim\,13000\times\,13000}$ for a $(k_x=0, k_y=0)$ state at ${B_z=20}$ T.
For AA-bt bilayer silicene, the model calculation takes into account the buckled honeycomb structure and complicated intra- and inter-layer atomic interactions. The combined effect of those ingredients and an external magnetic field is expected to generate the diverse physical phenomena, especially the magneto-electronic properties. \\

\section{Results and Discussion }

AA-bt bilayer silicene presents the unique and feature-rich energy dispersion, as demonstrated in Figs. 1(b) and 1(c) for 3D and 2D views, respectively. There are two pairs of valence and conduction bands due to the Si-$3p_z$-orbital $\pi$ bondings and they are slighly asymmetric about the Fermi level of $E_F=0$.
It shows the semimetallic behavior with zero gap and a finite density of states (DOS) at $E_F$, as clearly illustrated in Fig. 2(a).
The outer pair of conduction and valence energy bands is originated at higher and deeper energy ranges ($|E^c,v|\ge\,2$ eV). On the other hand, the pair of energy bands near $E_F$ is expected to dominate the low-energy essential properties of the system.
Interestingly, the stable and non-stable electronic valleys are formed from the electronic states near the high-symmetry points of the hexagonal first Brillouin zone. In particular, there exist the M, K and $\Gamma$ valleys with different conduction and valence band-edge state energies of about (0.5 eV, -0.51 eV), (1.19 eV, -1.21 eV), and (1.43 eV, -1.50 eV), respectively. These special valleys are expected to be closely related to the magnetic quantizations of the initial Landau levels, as discussed later. \\

The M valley belongs to the saddle point while the K and $\Gamma$ valleys correspond to the local extreme points, therefore they lead to the different van Hove singularities (Fig. 2(a)) and the diverse LL energy spectra. Furthermore the Dirac-cone energy dispersion is absent because the low-lying electronic states closest to the Fermi level are not formed at K/K$^\prime$ valleys. Instead of upward conduction and downward valence Dirac cone initiated from the K/K$^\prime$ points as in monolayer graphene \cite{18}, there exist the concave-downward conduction and concave-upward valence valley in AA-bt bilayer silicene. Such a significant difference between the two system might come from the more complicated and stronger atomic interactions of the latter.
It should be noticed that, the lowest conduction and highest valence electronic states in AA-bt stacking are located at the midway of M and $\Gamma$ points, moreover, they are separated by a very small energy spacing of $\sim\,8$ meV. As a result, the density of states near $E_F$ present the special structure with a finite value. Furthermore, the low-lying Landau levels do not correspond to the initial magneto-electronic states.
In general, the electronic properties in AA-bt bilayer silicene are in great contrast with those of monolayer silicene \cite{2} and a AA-stacked bilayer graphene \cite{19}. \\

\begin{figure}
\centering
{\includegraphics[width=0.8\linewidth]{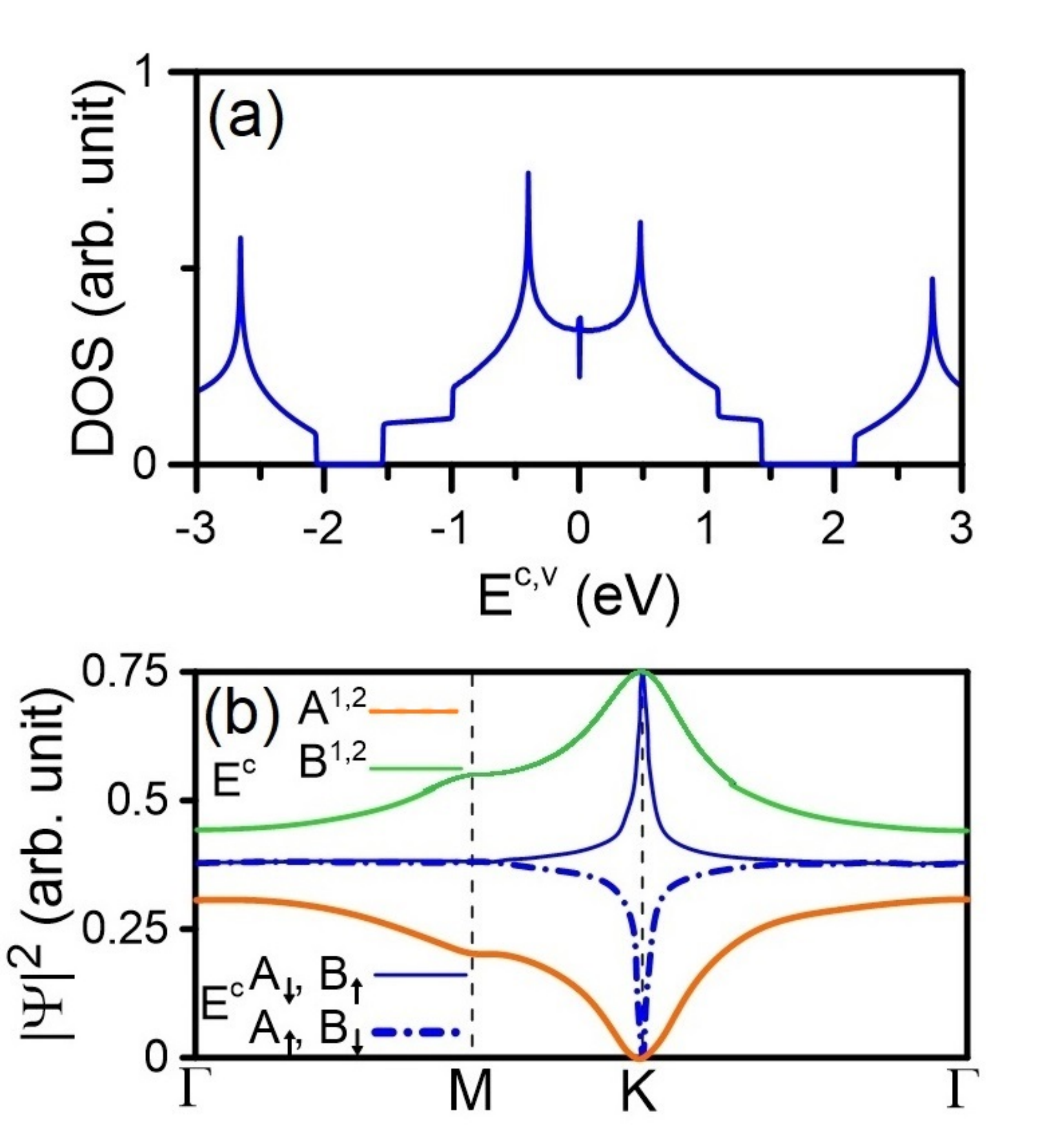}}
\caption{(Color online) The (a) van Hove singularities in density of states. The wave-vector-dependent wave functions are shown in (b). The solid and dashed blue curves of the A$^1$ and B$^1$ sublattices, respectively.}
\label{Figure 2}
\end{figure}

The van Hove singularities in the density of states are closely associated with three kinds of band-edge states, as illustrated in Fig. 2(a).
The DOS spectrum exhibit feature-rich structures, including the asymmetric peaks in the square-root divergent form, shoulder structures, and logarithmically divergent peaks.
There exist a pair of temple-like cusp structures crossing the Fermi level and they are separated by a quite small energy spacing of ${\sim\,8}$ meV.
These structures come from the extraordinary conduction and valence band structures near $E_F$ which could be considered as the one-dimensional parabolic dispersions.
Away from the Fermi level, the two pairs of prominent symmetric peaks arising from the 2D saddle point (M point) are located at (${0.50}$ eV $\&$ ${-0.52}$ eV) and (2.60 eV $\&$ ${-2.65}$ eV), respectively.
Regarding the extreme points, there appear the special shoulder structures at (1.10 eV, ${-1.15}$ eV) and (2.16 eV, ${-2.20}$ eV) for the K valley and (1.43 eV, ${-1.5}$ eV) for the $\Gamma$ one.
The special DOS spectrum in AA-bt bilayer silicene reflects the unique energy dispersion, it could be examined from the high-resolution STS experiments \cite{24, 25, 26, 27}. This measurement method is useful for the investigation of the interplay between the buckled structure and atomic interactions in bilayer silicene.
For AA-bt stacking, the DOS is relatively high near the M point compared with other high-symmetry points, leading to the very complicated LL energy spectrum, dissimilarly to those of monolayer graphene and silicene \cite{graphene, silicene}.
Roughly speaking, the M valley is regarded as the unstable one for the magnetic quantization where LLs could not be initiated from. \\

The Bloch wave functions consisting of the Si-${3p_z}$-orbital tight-binding functions on the four sublattices strongly depend on the wave vectors, as clearly indicated in Fig. 2(b). Those on the A and B sublattices are represented by the orange and green curves, respectively; the result for monolayer silicene is also shown as solid and dashed blue curves for comparison. The low-lying conduction band (and valence band; not shown) exhibit the equal distribution probability on the B$^1$ $\&$ B$^2$ [A$^1$ $\&$ A$^2$] sublattices because of the same ${(x,y)}$-plane projections with similar chemical environment.
Specifically, electronic states near the K point exhibit the vanishing A$^1$ and A$^2$ components and the identical B$^1$ and B$^2$ ones. Along the K${\rightarrow}$ M ${\rightarrow}$ $\Gamma$ directions, the B$^l$- and A$^l$-sublattice probabilities vary as ${0.5\rightarrow\,0.375\rightarrow\,0.265}$ and ${0\rightarrow\,0.125\rightarrow\,0.2355}$, respectively. Such behavior is quite different from those in monolayer silicene where the distribution of all sublatices become identical along M ${\rightarrow}$ $\Gamma$. Generally, the distribution probability of B$^l$ sublattices are evidently dominated the low-lying energy spectrum. The opposite is true for the higher conduction and deeper valence bands. That is to say, the the equivalence of the intralayer ${[A^l, B^l]}$ sublattices is thoroughly broken by the very strong interlayer hopping integrals (${\gamma_1}$ and $\gamma_2$).
On the other hand, monolayer silicene with significant spin-orbital coupling present similar distribution probabilities of the sublattices of the same spin state during the variation of wave vector due to the honeycomb lattice symmetry. \\

\begin{figure}
\centering
{\includegraphics[width=0.8\linewidth]{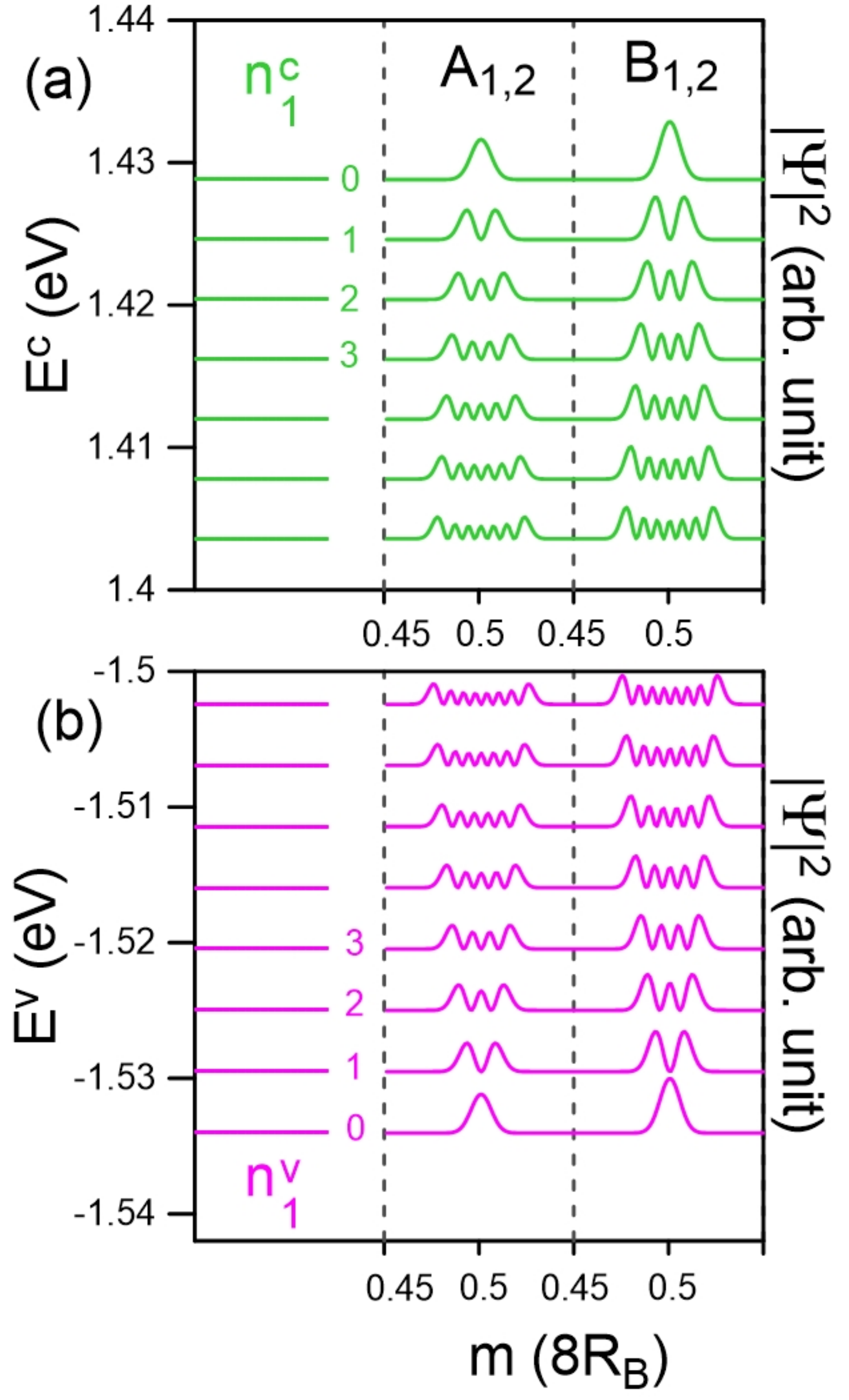}}
\caption{(Color online) The Landau levels and wave functions at $B_z$ = 40 T for the (a) conduction and (b) valence bands of the first group.}
\label{Figure 3}
\end{figure}

AA-bt stacking possesses the distinctive and diverse magnetic quantization phenomena, mainly owing to the feature-rich valley structures of the low-lying pair of conduction and valence bands. The LL energy spectra exhibit a number of interesting characteristics, including the LL state degeneracy, localization behavior of wave functions, sublattice dependence of localized oscillation modes, the well-behaved and non-well-behaved LLs, the complicated magnetic-field dependence, and the LL crossing phenomenon.
It is worth mentioning that the localization centers of LLs are continuously changed with the variation of ($k_x, k_y$), however, similar phenomena are observed for different electronic states.
Our numerical calculations in this work mainly focus on the magnetic quantization at ($k_x = 0, k_y = 0$) state, which is sufficient in understanding the essential magneto-electronic properties.
LLs could be classified into different groups based on the electronic valleys and the main features of spatial distributions. \\

As for LL initiated from the ${\Gamma}$-point top and bottom valley, the conduction and valence energy spectra present the explicit asymmetry behavior. The originated conduction and valence Landau levels are, respectively, located at ${1.43}$ eV and ${-1.53}$ eV for $B_z$ = 40 T, as shown in Figs. 3(a) and 3(b).
LLs exhibits a nearly uniform energy spectrum, as a result of the isotropically parabolic energy dispersion near the $\Gamma$ point, similarly to that of a 2D electron gas \cite{28}.
The well-behaved magnetic subenvelope functions on the four sublattices are localized at 0 and 1/2 of the $B_z$-enlarged unit cell, and they are degenerate.
The oscillation modes of LLs on all four subenvelope functions are equivalent.
Moreover, the LL wave functions on (A$^1$ $\&$ A$^2$) as well as (B$^1$ $\&$ B$^2$) sublattices are observed to be idential.
These special characteristics of LLs might be related to the equivalence of the intralayer [A$^l$, B$^l$] sublattices and the same chemical environment for the interlayer [A$^1$, A$^2$] $\&$ [B$^1$, B$^2$] sublattices.
The quantum number of each LL, $n_1^{c,v}$, is determined by the number of zero modes of its wave function on the dominated B$_i$ sublattices.
The fact that B$^l$ sublattices clearly dominate the energy spectra at ${\Gamma}$ valley is consistent with the zero-field wave-vector-dependent wave functions as shown in Fig. 2(b).
In general, each LL is four-fold degenerate by the spin and localization degrees of freedom. This is dissimilar to the eight-fold degeneracy of LLs in monolayer graphene and silicene \cite{graphene, silicene}. \\

\begin{figure}
\centering
{\includegraphics[width=0.8\linewidth]{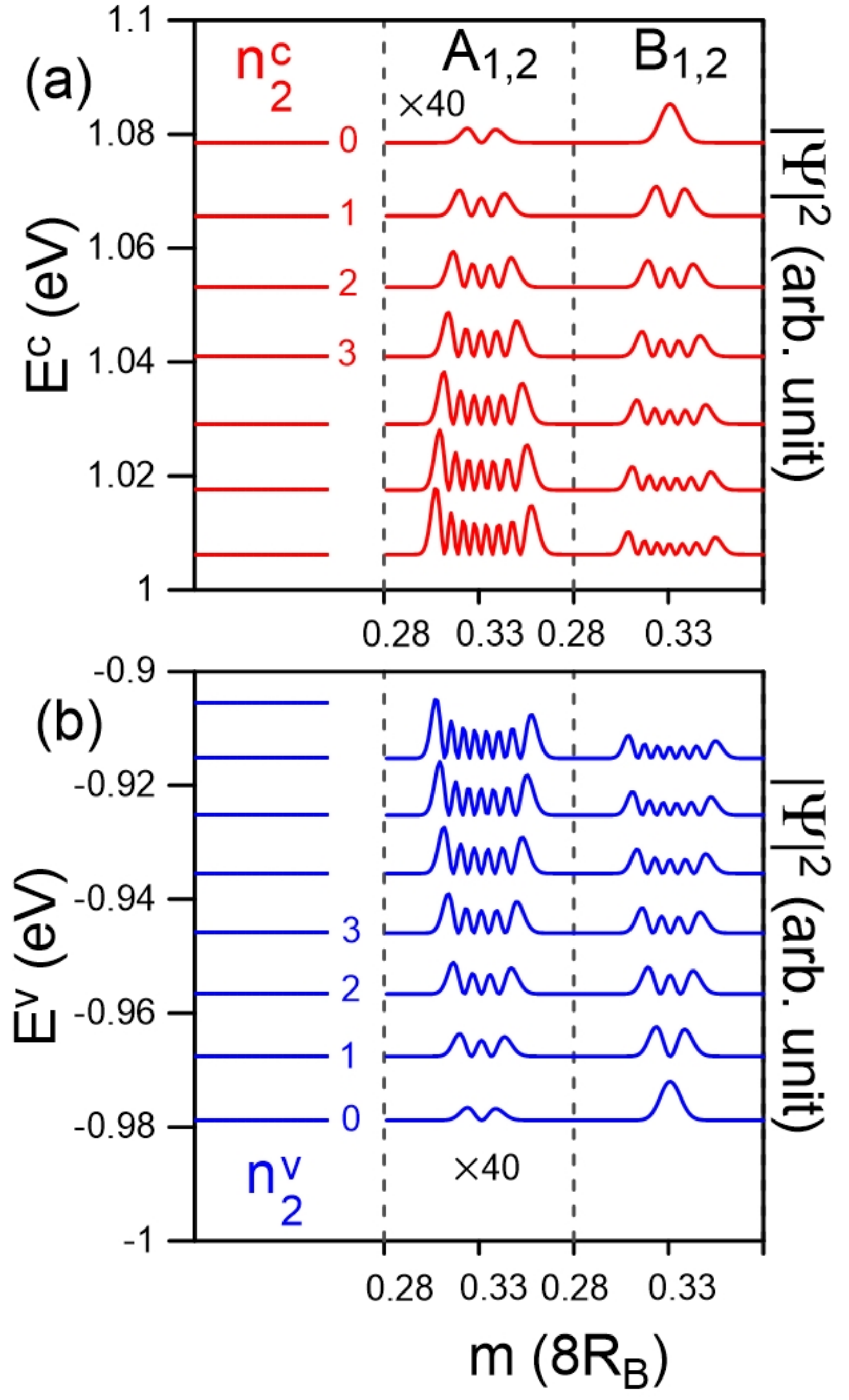}}
\caption{(Color online) The Landau levels and wave functions at $B_z$ = 40 T for the (a) conduction and (b) valence bands of the second group.}
\label{Figure 4}
\end{figure}

The conduction and valence Landau levels magnetically quantized from electronic states near the K/K$^\prime$ points display diverse characteristics. Here, there are LLs corresponding to both low-lying and outer pairs of energy bands, however, the former is the center of interest, as illustrated in Figs. 4(a) and 4(b).
Though the LL energy spectrum presents roughly uniform spacing, similarly to that near the ${\Gamma}$ valley, LLs show quite different behaviors.
The magnetic subenvelope functions are revealed at four distinct localization centers (1/6, 2/6 4/6 and 5/6) of the extended unit cell.
Especially, the LL spatial distributions are identical for 1/6 and 4/6 as well as 2/6 and 5/6 areas, leading to the eight-fold degeneracy of LLs, as observed in monolayer graphene \cite{6.22}.
The first few LLs are well-behaved and their quantum numbers ($n_2^{c,v}$) could be determined based on the wave functions of the evidently dominated $B^l$ sublattices.
The spatial distributions on the A$^l$ and B$^l$ sublattices exhibit one-mode difference, in which those on B$^l$ is higher (for 1/6 center) or less (for 2/6 center) than the corresponding ones on A$^l$, depending on the localization centers.
Apparently, the initial $n_2^{c,v} = 0$ LLs at 2/6 center are exceptionally four-fold degenerate because such magneto-electronic states only originate from the B$^l$ sublattices. \\

\begin{figure}
\centering
{\includegraphics[width=0.8\linewidth]{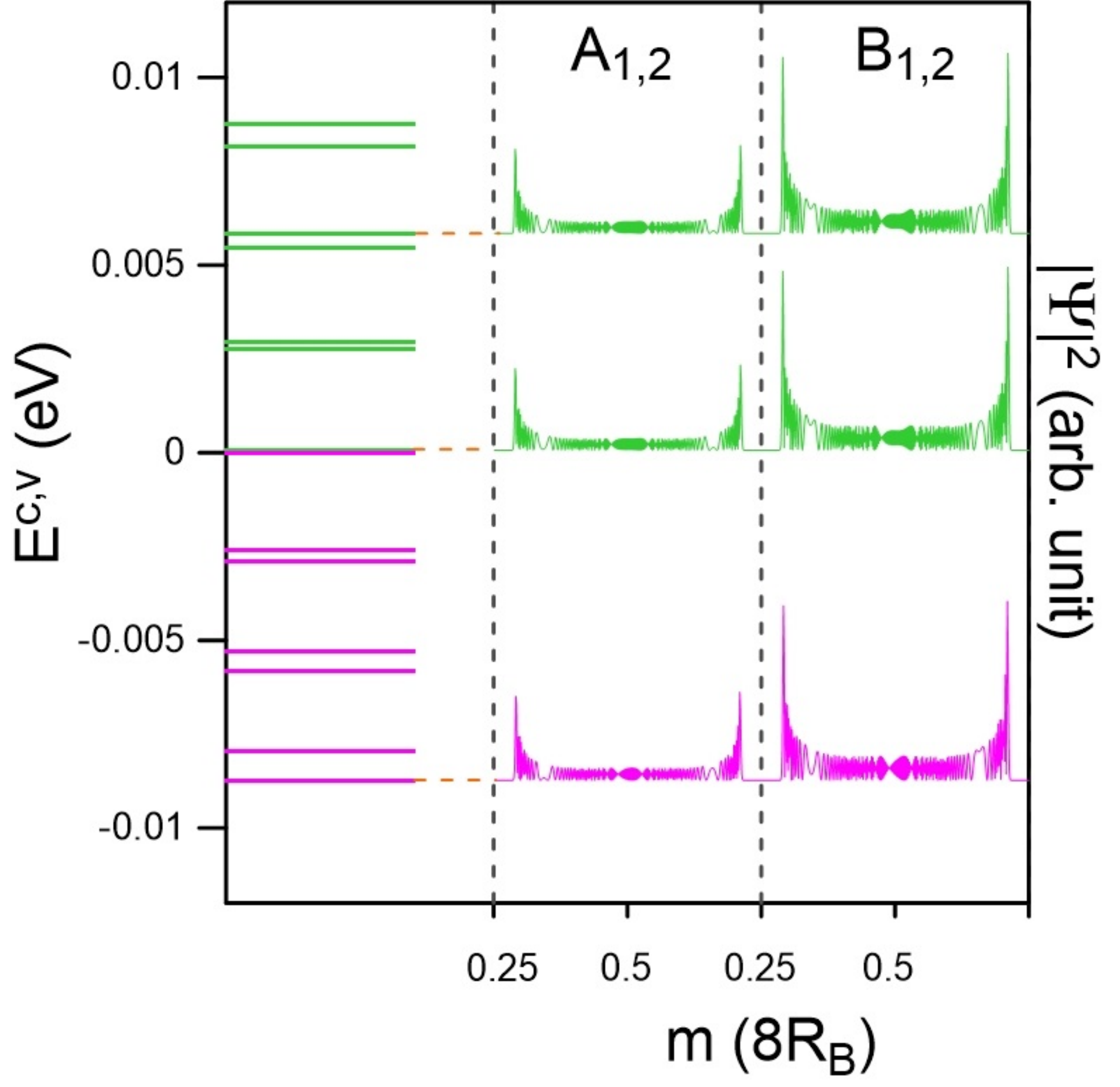}}
\caption{(Color online) The Landau levels and wave functions at $B_z$ = 40 T of the first group near the Fermi energy. The conduction and valence LLs are presented in green and pink, respectively.}
\label{Figure 5}
\end{figure}

Away from the initiated level, LLs quantized from $\Gamma$ and K valleys are slightly distorted because of the strong interlayer atomic interactions ($\gamma_1$ and $\gamma_2$), leading to the difficulty in defining their quantum numbers.
Regarding the $n_1^{c,v}$ LLs, though their oscillation distributions become much wider, which are ${\sim\,35\%}$ of a unit cell, they still remain two degenerate localization centers of 0 and 1/2.
This further illustrates the monotonous variation of the stable $\Gamma$ valley along the various directions.
For the $n_2^{c,v}$ LLs, the splitting of LL states at different localization centers of (1/6, 2/6, 4/6, and 5/6) is no longer observable.
Their quantum numbers are expected to be much smaller than those of the $n_1^{c,v}$ group due to the higher DOS in the K-related valleys (Fig. 2(a)). In addition, the magnetic quantization demonstrates the fact that the M-point saddle structure belongs to the K/K$^\prime$ valley. \\

In the vicinity of the Fermi level, there exist only conduction and valence LLs quantized from the $\Gamma$ valley, as clearly shown in Fig. 5.
Here, the magneto-electronic states are quite dense and complicated, therefore it is quite difficult to characterize the LLs.
The LL spatial distributions present thousands of oscillation modes which are localized at 0 and 1/2 centers.
Such unique magnetic quantization phenomenon is never observed in the other condensed-matter systems according to the previous theoretical and experimental studies \cite{13, graphene, silicene}.
Apparently, the high-resolution STS measurements \cite{24, 25, 26, 27} might not be able to directly identify the low-energy LLs in AA-bt bilayer silicene.
Nevertheless, the complex of low-energy magneto-electronic states in the AA-bt bilayer silicene should be associated with the other essential physical properties, such as, the delta-function-like van Hove singularities, magneto-optical absorption spectra with specific selection rules, quantum Hall transports, and inter-Landau-level damping and magnetoplasmon modes.
These should be worthy of a systematic investigation despite of possibly high barriers in the numerical technique and detailed analysis. \\

\begin{figure}
\centering
{\includegraphics[width=0.8\linewidth]{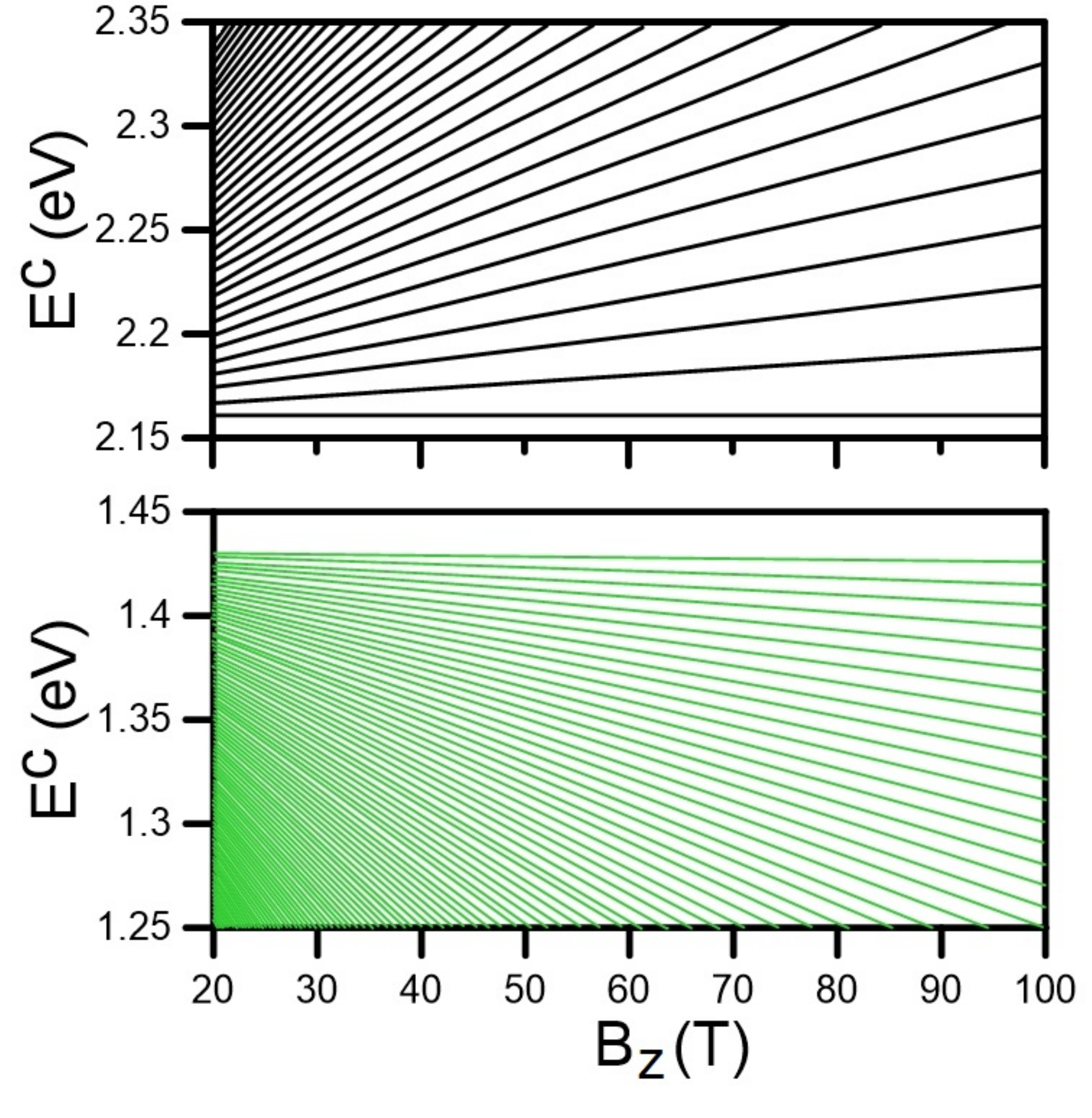}}
\caption{(Color online) The magnetic-field-dependent conduction Landau level energy spectra for (a) the first (green curves) and (b) third (black curves) groups.}
\label{Figure 6}
\end{figure}

The magnetic-field-dependent energy spectra are critical for the particularized comprehension of the magnetic quantization phenomena.
Though the conduction and valence LLs exhibit asymmetric spectra, their main features are equivalence, therefore the following discussion will focus on the conduction bands.
The quantized LLs of the outer pair of energy bands show the linear $B_z$-dependence, as demonstrated in Fig. 6(a) for the conduction spectrum.
The split energy of the neighboring LLs gradually increases with the growth of field strength.
These quantization characteristics indicate the stable K/K$^\prime$ valleys for such electronic structure.
On the other hand, the dependence of LL groups corresponding to the low-lying energy bands on the variation of magnetic field is much more complicated.
The energy spectra are the combination of both $n_1^{c,v}$ and $n_2^{c,v}$ LL groups with frequent intragroup and intergroup crossing behaviors, as illustrated in Figs. 6(b), 7 and 8. \\

\begin{figure}
\centering
{\includegraphics[width=0.8\linewidth]{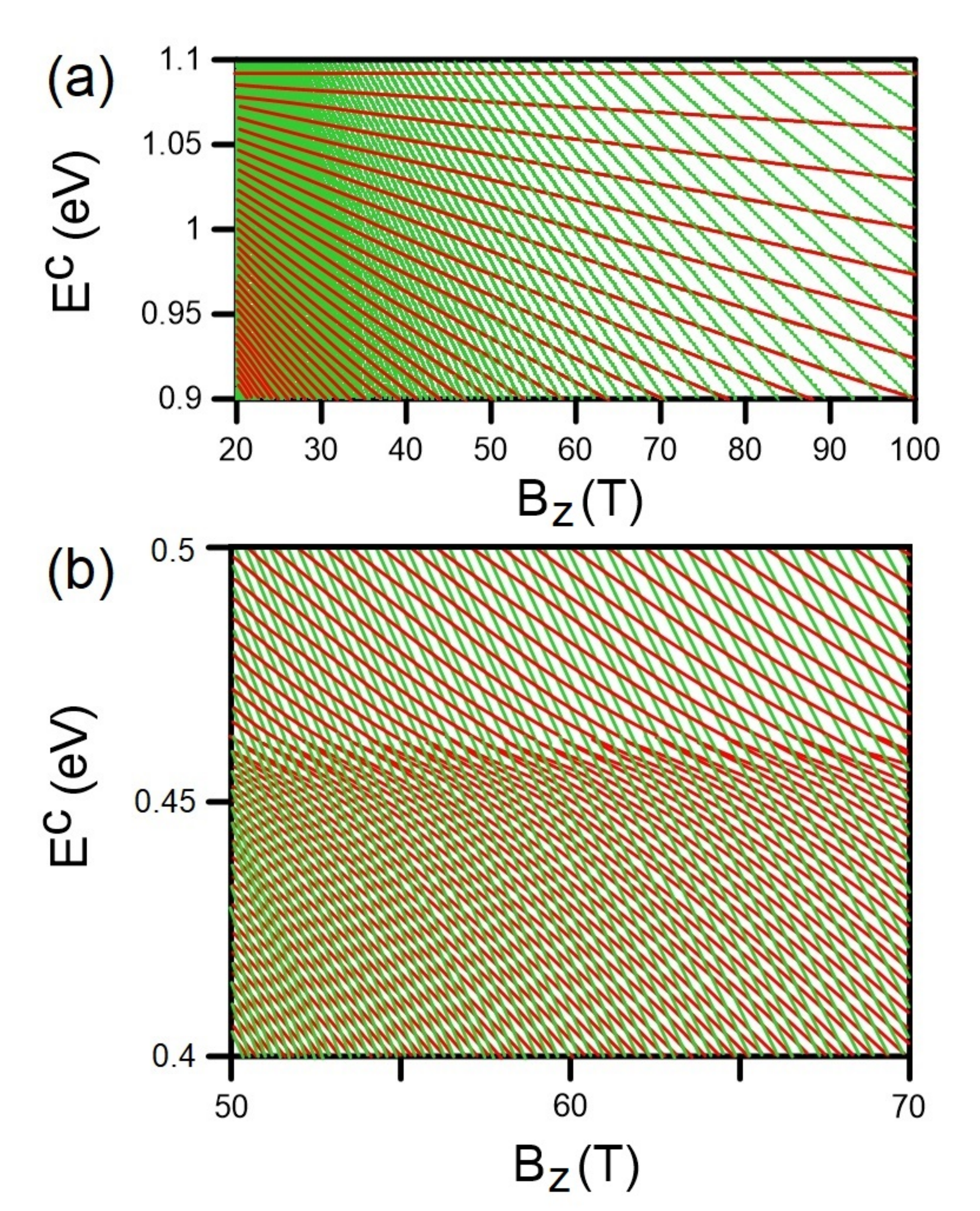}}
\caption{(Color online) The magnetic-field-dependent conduction Landau level energy spectra for the first (green curves) and second (red curves) groups at (a) higher and (b) lower energy ranges.}
\label{Figure 7}
\end{figure}

Concerning the $n_1^{c}$ LLs originated from the $\Gamma$ point, the initiated LLs are linearly dependent on $B_z$ without any intragroup crossing (Fig. 6(b)), similarly to that of the outer energy bands but with smaller LL energy spacing.
With decreasing energy, the $B_z$-dependent spectrum contain of both the $n_1^{c}$ (green curves) and $n_2^{c}$ (red curves) LLs, as shown in Figs. 7(a) and 7(b).
Interestingly, the two groups of LLs present crossing phenomenon without any hybridization of magneto-electronic states. This is because the two stable valleys of K and ${\Gamma}$ are independent of each other.
The evolution of $n_2^{c}$ LLs with the increase of magnetic field strength have simply linear behavior for sufficiently high energy, $E^c \geq 0.5$ eV, as clearly presented in Fig. 7(a).
On the contrary, LLs at $E^c \sim 0.45$ eV are formed in the shoulder-like structure, corresponding the energy dispersion around the M point.
The main reason for this is that the electronic states at very high DOS can not be quantized into the well-behaved LLs. \\

The magneto-electronic properties of AA-bt bilayer system are in great contrast with those in monolayer silicence and graphene \cite{graphene, silicene}.
While the former shows the initial LLs at higher and lower energy ranges, those of the latter begin near $E_F=0$.
Apparently, the main differences lie on the main characteristics of LLs around the Fermi level which are very complicated for AA-bt stacking.
Monolayer systems possess the conduction and valence Dirac cones with the Dirac point or extremely narrow gap at the K/K$^\prime$ point. The low-lying LLs present the well-behaved spatial distributions and therefore their quantum numbers are easily determined based on the number of zero modes.
Moreover, the $B_z$ dependence of LL energies exhibit linear form in the absence of crossing phenomenon, dissimilarly to the frequent crossing LLs in AA-bt stacking.
The above-mentioned divergent points of AA-bt bilayer silicene compared with other monolayer systems mainly come from its unique intrinsic properties, such as the buckling structure with opposite ordering and significant interlayer atomic interactions. \\

\section{Concluding Remarks}

We have presented a thorough investigation of the unique and diverse magnetic quantization phenomena in AA-bt bilayer silicene using a generalized tight-binding model.
This material possesses the special electronic structure with two pairs of energy bands, in which the low-lying pair shows interesting oscillation shape.
Remarkably, the lowest conduction and highest valence states near the Fermi level are away from the K point, dissimilarly to that of graphene and other 2D systems.
The complicated energy dispersion is closely related to the feature-rich magnetically quantized Landau levels.
LLs originated from K and $\Gamma$ valleys are quite different in the main characteristics, covering the LL degeneracy, the spatial distributions, the dominated sublattices, and the magnetic-field dependence.
Especially in the vicinity of Fermi energy, there are so many magneto-electronic states and they do not present simple wave function modes.
The geometry symmetry, intralayer and interlayer atomic interactions, and effect of a perpendicular magnetic field are responsible for the peculiar LL energy spectra in AA-bt bilayer silicene. \\

\begin{acknowledgements}
We would like to acknowledge the financial support from the Ministry of Science and Technology of Taiwan (R.O.C.) under Grant No. MOST 105-2112-M-017-002-MY2. D.H. would like to thank the supports from the AFOSR and from the DoD Lab-University Collaborative Initiative (LUCI) Program. G.G. would like to acknowledge the support from the Air Force Research Laboratory (AFRL) through Grant \#12530960.
\end{acknowledgements}

\newpage

$\textbf{References}$

\end{document}